\documentclass[12pt]{article}
 \usepackage{amsmath}
  \usepackage[dvips]{graphicx}

\newcommand{\nc}{\newcommand}
\nc{\non}{\nonumber}
\nc{\pt}{p_{{}_T}}
\nc{\lmc}{\Lambda_c^+}
\nc{\plmc}{\vec{\Lambda}_c^+}
\nc{\all}{A_{LL}}
\nc{\dg}{\Delta G(x,Q^2)}
\nc{\stil}{\tilde{s}}
\nc{\ttil}{\tilde{t}}
\nc{\util}{\tilde{u}}
\nc{\shs}{\hat{s}}
\nc{\ths}{\hat{t}_1}
\nc{\uhs}{\hat{u}_1}
\nc{\cosec}{\rm cosec}
\nc{\mpr}{m_p}
\nc{\mc}{m_c}
\nc{\mlc}{m_{\lmc}}
\nc{\pa}{p_{{}_a}}
\nc{\pb}{p_{{}_b}}
\nc{\pA}{p_{{}_A}}
\nc{\pB}{p_{{}_B}}
\nc{\plc}{p_{\lmc}}
\nc{\pc}{p_{{}_c}}
\begin{document}
\pagestyle{empty} \setlength{\footskip}{2.0cm}
\setlength{\oddsidemargin}{0.5cm} \setlength{\evensidemargin}{0.5cm}
\renewcommand{\thepage}{-- \arabic{page} --}
   \def\thebibliography#1{\centerline{\large \bf REFERENCES}
     \list{[\arabic{enumi}]}{\settowidth\labelwidth{[#1]}\leftmargin
     \labelwidth\advance\leftmargin\labelsep\usecounter{enumi}}
     \def\newblock{\hskip .11em plus .33em minus -.07em}\sloppy
     \clubpenalty4000\widowpenalty4000\sfcode`\.=1000\relax}\let
     \endthebibliography=\endlist
   \def\sec#1{\addtocounter{section}{1}\section*{\hspace*{-0.72cm}
     \normalsize\bf\arabic{section}.$\;$#1}\vspace*{-0.3cm}}
\begin{flushright}
KOBE-FHD-02-05\\
YNU-HEPTH-02-103\\
(hep-ph/0211046)
\end{flushright}
\renewcommand{\thefootnote}{$\dag$}
\begin{center}
{\bf  \Huge Charmed Hadron Production in}\vspace*{0.5cm}\\
{\bf \Huge Polarized \mbox{\boldmath{$pp$}} 
Collisions}{\Large \footnote{Talk presented by K. Ohkuma at
the 15th International Spin Physics Symposium, \\
\phantom{aaaa}BNL,
Sep.  9-14, 2002}}
\vspace*{1cm}\\

{\sc \Large Toshiyuki MORII}\vspace*{0.2cm}\\
Division of Sciences for Natural Environment,\\
Faculty of Human Development,\\
Kobe University, Nada, Kobe 657-8501, JAPAN\\
Electronic address{\tt :morii@kobe-u.ac.jp}\vspace{1.0cm}\\

{\sc \Large Kazumasa OHKUMA}\vspace*{0.2cm}\\
Department of Physics,\\
Faculty of Engineering,\\
Yokohama National University,\\
 Hodogaya, Yokohama
240-8501, JAPAN\\
Electronic address{\tt :ohkuma@phys.ynu.ac.jp}
\end{center}
\vspace*{0.5cm}

\centerline{\bf \large ABSTRACT}
\vspace*{0.2cm}
\baselineskip=20pt plus 0.1pt minus 0.1pt

To extract information about polarized gluons in the proton,
production of charmed hadrons, in particular,
$\Lambda_c^+$ baryon in $pp$ collisions was studied. 
We calculated the transverse momentum distribution
and the pseudo-rapidity  distribution of the 
spin correlation asymmetry $A_{LL}$ between the initial 
proton and the produced $\Lambda_c^+$. 
Those statistical sensitivities were also 
calculated under the condition of RHIC experiment.
We found that the pseudo-rapidity distribution of $A_{LL}$
is promising for testing the model of polarized gluons in the proton
and also the spin-dependent fragmentation model of a charm quark
decaying into $\Lambda_c^+$ baryon.

\newpage

\renewcommand{\thefootnote}{$\sharp$\arabic{footnote}}
\pagestyle{plain} \setcounter{footnote}{0}
\pagestyle{plain} \setcounter{page}{1}
\baselineskip=18.0pt plus 0.2pt minus 0.1pt
\section{Introduction}

~~The Relativistic Heavy Ion Collider (RHIC) at Brookhaven
National Laboratory has just started to explore the internal 
structure of proton. 
One of the important purposes of those RHIC experiments is to study 
the behavior of polarized gluons in the proton.  As is well known,
the proton spin is given by the sum of the spin carried by quarks 
$\Delta\Sigma$ and gluons $\Delta G$, and their orbital 
angular momenta $\langle L_Z\rangle$.  
In these years, a great deal of efforts have been made for 
extracting those components from polarized structure functions
of nucleons\cite{rpsp}.  Based on the next--to--leading order QCD analyses
on the polarized structure functions $g_1(x)$, the contribution of 
quarks to the proton spin is well known.  However,  
knowledge on polarized gluons in the proton is still poor.
To understand the origin of the nucleon spin, it is very important
to know how gluons polarize in the nucleon.  So far, several interesting
processes have been proposed for extracting $\Delta G$.  Here we 
also propose a different process to obtain more detailed 
information on polarized gluons, expecting the forthcoming RHIC
experiment.  The processes which we propose here are the polarized
charmed hadron production, i.e. $p\vec{p}\to \vec{\Lambda}_c^+X$ and 
$p\vec{p}\to \vec{D}^*X$, in the polarized proton--unpolarized proton
collision,\footnote{Though we have calculated even for $D^*$ production,
we focus only on the $\Lambda_c^+$ production in this report, because the 
main point of the result remain unchanged.}
which will be observed at the RHIC experiment.
We study which observables are useful for
extracting information about polarized gluons in the proton and also
discuss its sensitivity.

\section{$\lmc$ Baryon Production in Proton--Proton Collision}

~~In the process on which we focus here, the $\Lambda_c^+$ baryon 
is expected to have some advantageous properties 
for probing behavior of polarized gluons in the proton.  
Those properties are as follows;
%
%
\begin{enumerate}
\item A charm quark which is one of constituents of the $\Lambda_c^+$ baryon
is dominantly produced via gluon-gluon fusion in $pp$ reaction,
because charm quarks
are tiny contents in the proton. 
Thus, the cross section of this process is directly 
proportional to the gluon distribution in the proton.
\item Since the $\Lambda_c^+$ baryon is composed of
a charm quark and  
antisymmetrically combined light up and down quarks,
the spin of $\Lambda_c^+$ baryon is expected to be almost equal to 
the spin of its constituent charm quark.   
\item Since a charm quark is heavy, it must be very rare 
for the charm quark to change its spin
arrangement during its fragmentation into a $\Lambda_c^+$ baryon. 
In other words, the spin direction of the charm quark produced in 
the subprocess is 
expected to be kept in the $\Lambda_c^+$ baryon produced 
in the final state. 
\end{enumerate}
%
After all, the spin of the $\lmc$ is in strong correlation to
the polarization of gluons in the proton.
Therefore, by observing the spin correlation between the 
polarized proton in the initial state and the 
polarized $\lmc$ in the final state, we can get, rather clearly,
information on the polarized gluon in the proton.

\section{Spin Correlation Asymmetry and its Statistical Sensitivity}

~~To study the polarized gluon distribution in the proton, we introduced
the spin correlation asymmetry of the target 
polarized-proton and
produced $\Lambda_c^+$ baryon~\cite{osm};
\begin{eqnarray}
A_{LL}&=&\frac{d \sigma_{++} - d\sigma_{+-} + d \sigma_{--}- d\sigma_{-+}}
{d \sigma_{++} + d \sigma_{+-}+d \sigma_{--} + d \sigma_{-+}}\nonumber\\
&\equiv&\frac{{d \Delta \sigma}/{d X}}
{{{d\sigma}/{d X}}},~~(X=\pt~{\rm or}~\eta), \label{eq:all}
\label{all}
\end{eqnarray}
where $d \sigma_{+ -}$, for example, denotes the spin-dependent
differential cross section with the positive helicity of the target proton
and the negative helicity of the produced $\Lambda_c^+$ baryon.
$\pt$ and $\eta$, which are represented by $X$ in Eq.(\ref{all}),
are transverse momentum and pseudo-rapidity of
produced $\lmc$, respectively. 

According to the quark-parton model,
$d \Delta \sigma / d X$ can be expressed as
\begin{eqnarray}
&&\frac{d \Delta \sigma}{d X}
 = \int^{Y{{}^{\rm max}}}_{Y{{}^{\rm min}}}
\int^{1}_{x^{{}^{\rm min}}_{{}_a}} 
\int^{1}_{x^{{}^{\rm min}}_{{}_b}}
          G_{p_{{}_A}\rightarrow g_{{}_a}}(x_a,Q^2)
   \Delta G_{\vec{p}_{{}_B} \rightarrow \vec{g}_{{}_b}}(x_b,Q^2)
   \Delta {D}_{\vec{c}\rightarrow \vec{\Lambda}_c^+}(z)\non \\
&&\phantom{\frac{d \Delta \sigma (\lambda,h)}{d \pt}=}
\times \frac{d \Delta \hat{\sigma}}{d \hat{t}}
J
dx_a dx_b dY,~~~\left(X,Y = \eta~{\rm or}~\pt ~(X \neq Y)\right),
\label{dcross}
\end{eqnarray}
with
$$
J\equiv \frac{2 s \beta \pt^2{\rm cosh}\eta}{z \hat{s}\sqrt{m_c^2+
\pt^2{\rm cosh^2 \eta}}}, ~~\beta\equiv\sqrt{1-\frac{4 m_p^2}{s}}
$$
where 
$ G_{p_{{}_A} \rightarrow g_a}(x_a,Q^2)$,
$\Delta G_{\vec{p}_{{}_B} \rightarrow \vec{g}_b}(x_b,Q^2)$
and $\Delta {D}_{\vec{c} \rightarrow \vec{\Lambda}_c^+}(z)$  
represent the unpolarized gluon distribution function, the polarized gluon
distribution function and the spin-dependent fragmentation function of the
outgoing charm quark decaying into a polarized $\plmc$, respectively.
 $d \Delta \hat{\sigma}/d \hat{t}$ is the spin-dependent
differential cross section of the subprocess 
and  $J$ is the Jacobian which transforms the variables  $z$  and $\hat{t}$ 
into $\pt$ and $\eta$.
In the expression of Eq.(\ref{dcross}),  $\pt$ and $\eta$ are described
as $X$ or $Y$.

Statistical sensitivities of $\all$ for the $\pt$  and  
$\eta$ distribution are estimated by using the following formula;
\begin{equation}
\delta \all \simeq \frac{1}{P}\frac{1}{
\sqrt{b_{\Lambda_c^+}~\epsilon~L~T~\sigma}}.
\end{equation}
To numerically estimate the value of $\delta \all$, here 
we use following parameters:
operating time; $T=$100-day, the beam polarization; $P=$70\%,
a luminosity; $L=8\times 10^{31}~(2\times 10^{32}$)~cm$^{-2}$ sec$^{-1}$ for
$\sqrt{s}=200~(500)~$ GeV,
the trigger efficiency; $\epsilon =10\%$ for detecting produced
$\Lambda_c^+$ events and a branching ratio; $b_{\Lambda_c^+}\equiv 
{\rm Br}(\lmc \to p K^- \pi^+)\simeq5\%$~\cite{pdg}.
The purely charged decay mode is needed to measure the polarization of
produced $\lmc$. 
$\sigma$ denotes the unpolarized cross section integrated over suitable
$\pt$ or $\eta$ region.
%
%
\section{Numerical Calculations}
~~To carry out the numerical calculation of $A_{LL}$, 
we used, as input parameters,
$m_c = 1.20$ GeV, $m_p = 0.938$ GeV and $\mlc = 2.28$ GeV\cite{pdg}.
We limited the integration region of $\eta$ and 
$\pt$ of produced $\lmc$  as
$-1.3 \leq \eta \leq 1.3$ and
3 GeV $\leq \pt \leq$ 15(40) GeV, respectively, 
for  $\sqrt{s}=200(500)$ GeV.  The range 
of $\eta$ and the lower limit of $\pt$ were selected 
in order to get rid of the contribution from
the diffractive $\lmc$ production. 
As for the upper limit of $\pt$, we took it as described above, 
for simplicity, though the kinematical maximum of $\pt$
of produced $\Lambda_c^+$ is 
slightly larger than 15 GeV and 40 Gev for $\sqrt{s}=$200 GeV 
and 500 GeV, respectively.
In addition, we took the AAC\cite{aac} and  GRSV01~\cite{grsv}
parameterization models for the polarized gluon distribution 
function and the GRV98~\cite{grv} model for the
unpolarized one.
Though both of AAC and GRSV01 models excellently reproduce the experimental 
data on the polarized structure function of nucleons $g_1(x)$,
the polarized gluon distributions for those models are quite different. 
In other words, the data on polarized structure function of nucleons 
$g_1(x)$ alone are not enough to distinguish the model of gluon distributions.
Since the process is semi-inclusive, the fragmentation function of a 
charm quark to $\Lambda_c^+$ is necessary to carry out numerical calculations.
For the unpolarized fragmentation function, 
we used Peterson fragmentation function, 
$D_{c \to \Lambda_c^+}(z)$~\cite{pdg,peter}.
However, since we have no data, at present, 
about polarized fragmentation
functions for the polarized $\Lambda_c^+$ production, we took the
following ansatz for the polarized fragmentation function
$\Delta D_{\vec{c}\to \vec{\Lambda}_c^+}(x)$,
\begin{equation}
\Delta D_{\vec{c}\to \vec{\Lambda}_c^+}(z)= C_{c\to \Lambda_c^+} 
D_{c \to \Lambda_c^+},
\end{equation}
where $C_{c \to \Lambda_c^+}$ is a 
scale-independent spin transfer coefficient.
In this analysis, we studied two cases: (A) 
$C_{c\to \Lambda_c^+}=1$ (non-relativistic quark model) and 
(B) $C_{c\to \Lambda_c^+}=z$ (Jet fragmentation model~\cite{jet}).
As we discussed before,
if the spin of $\lmc$ is same as the spin of charm quark produced in
subprocess, the model (A) might be a reasonable scenario.

\begin{figure}[htb]
\includegraphics[scale=0.31]{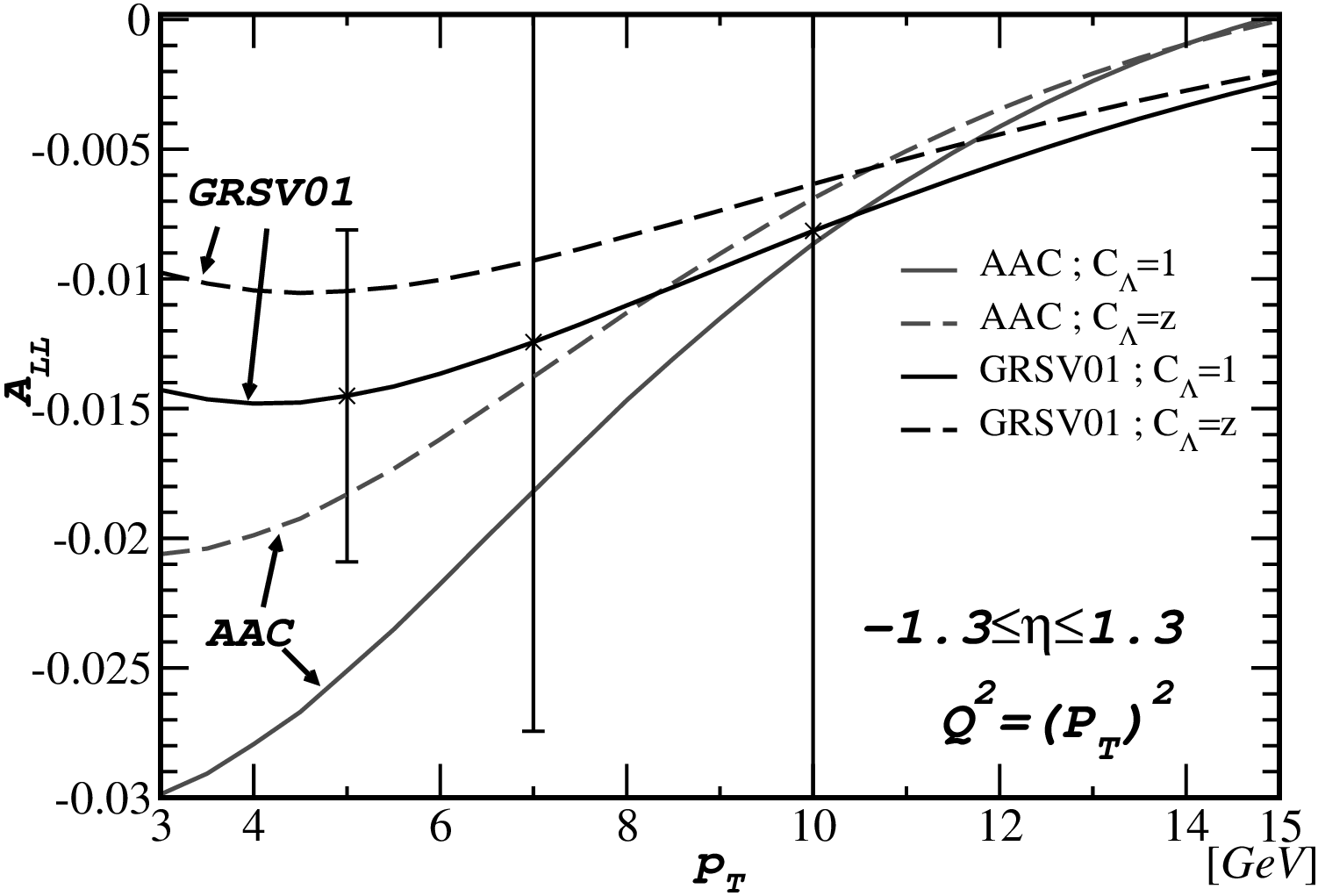}
\includegraphics[scale=0.31]{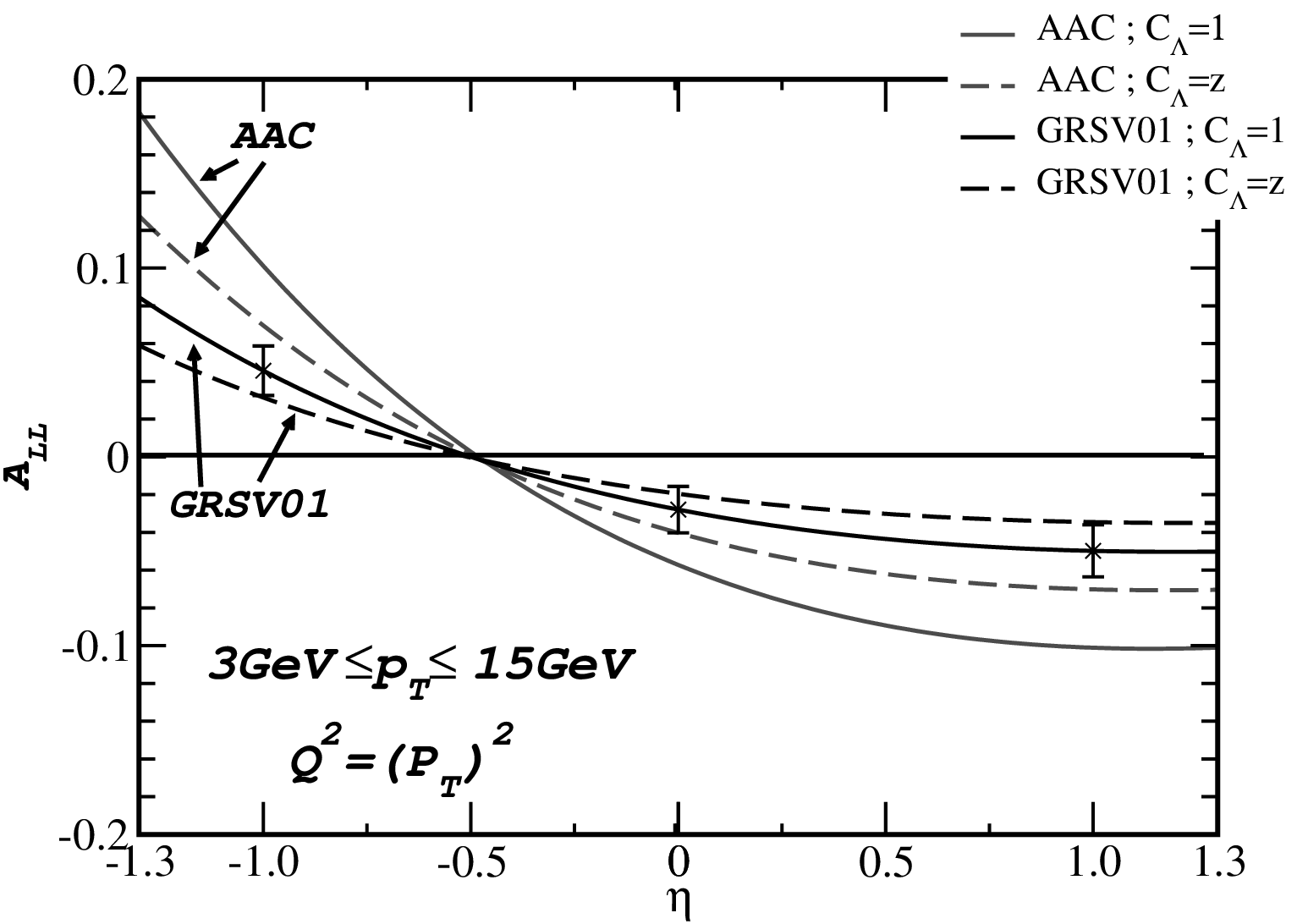}
  \caption{$\all$ as a function of $\pt$(left panel) 
   and $\eta$ (right panel) at $\sqrt{s}=$200 GeV}
\label{200}
\end{figure}

\begin{figure}[htb]
\vspace*{0.5cm}
  \includegraphics[scale=0.31]{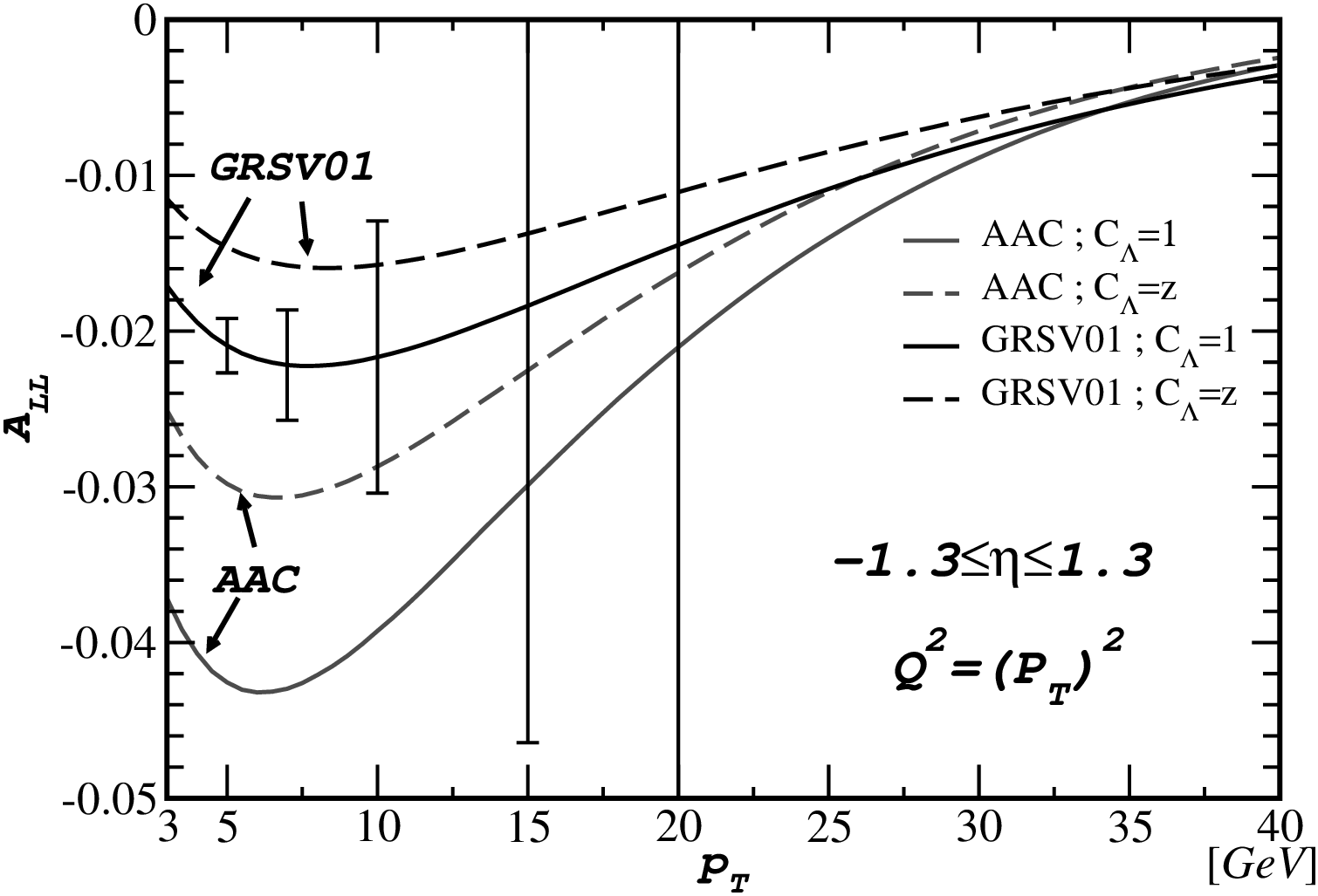}~
  \includegraphics[scale=0.31]{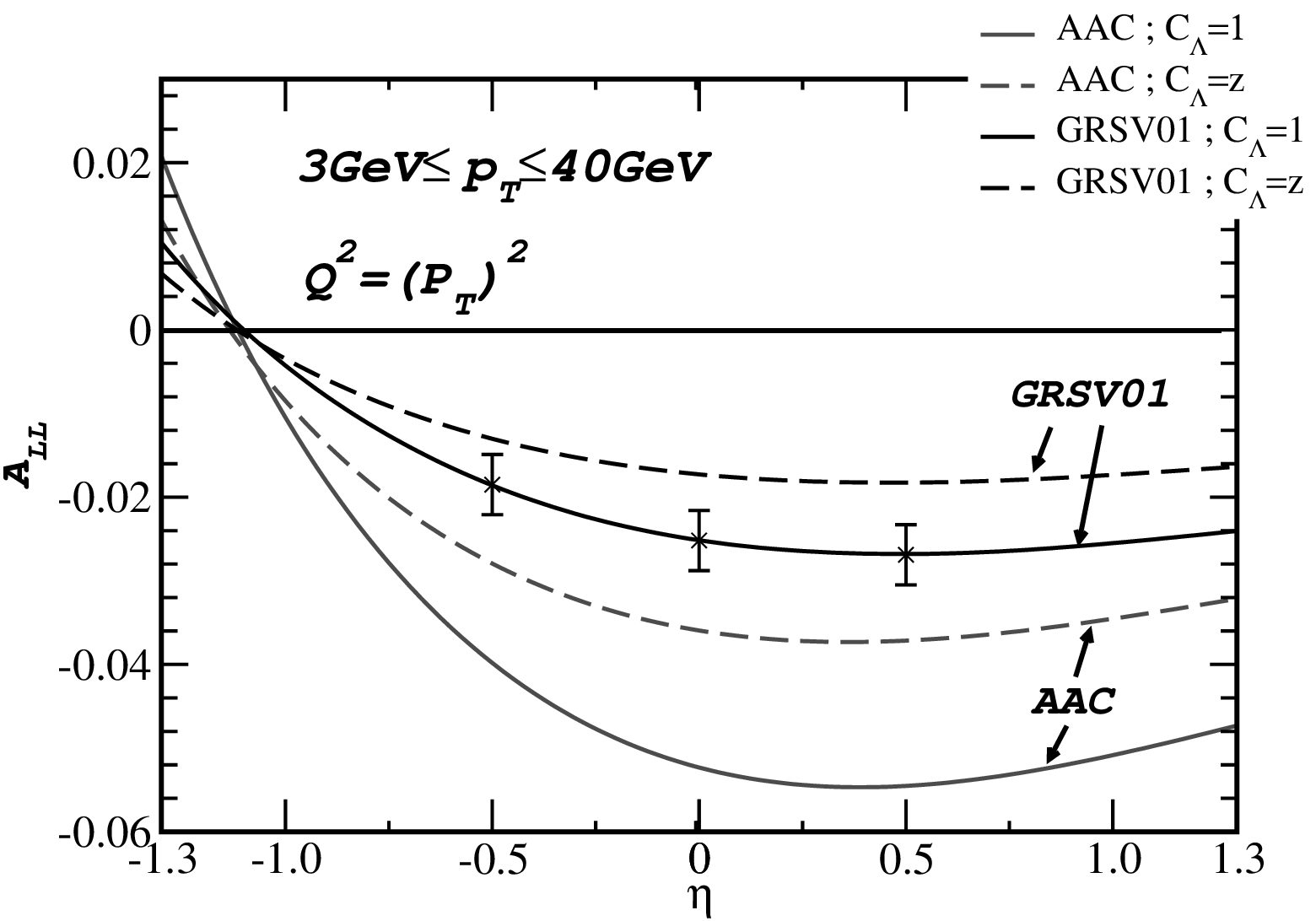}
  \caption{The same as in Fig.~{\ref{200}}, but for $\sqrt{s}=$500 GeV}
 \label{500}
\end{figure}

Numerical results of $\all$ are shown in Fig.~\ref{200} 
and Fig.~\ref{500}. In those figures,
statistical sensitivities, $\delta \all$, are also attached 
to the solid line of $\all$ which 
is calculated for the case of the GRSV01 parametrization model of 
polarized gluon and the non-relativistic fragmentation model.
\footnote{Note that as shown from Eq.(4), $\delta A_{LL}$ 
does not depend on both of the model of polarized gluons and the model 
of fragmentation functions.} 
From these results, we see that
the $\eta$ distributions of $\all$ are more effective observables
than the $\pt$ distributions at $\sqrt{s}=$200 GeV and 500 GeV.
As shown in the right panel of Fig.~\ref{500} given at $\sqrt{s}=$500 GeV,
we could distinguish the parametrization models of 
polarized gluon as well as the models of the spin-dependent 
fragmentation function though the magnitude of $\all$ is rather small.
At $\sqrt{s}=$200 GeV, the magnitude of $\all$ for $\eta$ distribution 
becomes larger, though statistical sensitivities
are not so small.
If the integrated luminosity at  $\sqrt{s}=$200 GeV could be
large and the detection efficiency, $\epsilon$, is improved,
this observable could be promising to distinguish
not only the models of  $\Delta G(x)$ but also
the models of $\Delta D(z)$.
For the $\pt$ distribution of $\all$,
$\delta \all$ become rapidly larger with increasing $\pt$ and
we cannot say anything from those region.
However, if we confine the kinematical region in 
rather small $\pt$ range like
$\pt=3\sim$5(10) GeV at $\sqrt{s}=200(500)$ GeV,  
it might be still effective.
\\

\section{Concluding Remark}
~~~To extract information on the polarized gluon distribution 
in the proton, the charmed hadron production processes at RHIC 
experiments have been proposed.
(Actually, only $\lmc$ process was discussed in this report.)
The spin correlation asymmetry $\all$ between the initial proton 
and the produced $\lmc$ was calculated for $\pt$ and $\eta$ distributions 
with statistical sensitivities which were estimated using RHIC parameters. 
We found that $\all$ is rather sensitive to the model 
of $\Delta G(x)$ and $\Delta D(z)$.
The $\eta$  distribution of $\all$ could be promising for 
distinguishing the parametrization model of polarized gluons
as well as the model of spin-dependent fragmentation of a charm quark
into $\lmc$. 
%
%
\section*{Acknowledgments}
One of the authors, (K.O), would like to thank the organizers and fellowship
committee of SPIN2002 for giving a chance to present this contribution
at the symposium.\vspace*{1cm}\\
%

\end{document}